\documentclass[12pt,preprint]{aastex}
 \def\ltsima{$\; \buildrel < \over \sim
\;$} \def\simlt{\lower.5ex\hbox{\ltsima}}            
\def\gtsima{$\; \buildrel > \over \sim \;$}
\def\simgt{\lower.5ex\hbox{\gtsima}}            

\begin{document} 

\title{\sc Opacity Variations in the Ionized Absorption in NGC
3783: A Compact Absorber}
\author{Y. Krongold\altaffilmark{1}, F. Nicastro\altaffilmark{1},
 N.S. Brickhouse\altaffilmark{1}, M. Elvis\altaffilmark{1}, S. Mathur
 \altaffilmark{2} }

\altaffiltext{1}{Harvard-Smithsonian Center for Astrophysics. 60
Garden Street, Cambridge MA 02138}
%

\altaffiltext{2}{Department of
Astronomy, Ohio State University, 140 West 18th Avenue, Columbus,
OH 43210}


\begin{abstract}

We show that the  Fe ({\sc vii-xii}) M-shell unresolved
transition array (UTA) in the NGC 3783 900 ks {\em Chandra} HETGS
observation clearly changes in opacity in a timescale of 31 days
responding to a factor of $\sim 2$ change in the ionizing continuum.
The opacity variation is observed at a level $>10\sigma$.
There is also evidence for variability in the O {\sc vi} K edge (at
$\sim3\sigma$). The observed changes are consistent with the gas
producing these absorption features (the low ionization component)
being  close to photoionization equilibrium. The gas
responsible for the Fe ({\sc xvii-xxii}) L-shell absorption (the high
ionization component), does not seem to be responding as expected in
photoionization equilibrium. 
The observed change in opacity for the UTA implies a density $>1\times
10^{4}$  cm$^{-3}$, and so locates the gas within 6 pc of
the X-ray source. The scenario in which
the gas is composed of a continuous radial range of ionization
structures is ruled out, as in such scenario, no opacity variations are
expected. Rather, the structure of the absorber is likely composed by
heavily clumped gas.

\end{abstract}

\section{Introduction}

The location of the wind seen to be outflowing from at least half of all Active
Galactic Nuclei (AGNs) in X-ray and UV spectra (Crenshaw, Kraemer and
George 2003) is
a matter of importance to understanding the structure of these objects, but is
currently a matter of great debate. If the wind originates from the accretion
disk, widely thought to be the origin of the AGN continuum, then all the
emission and absorption features in AGN spectra, including the broad emission
lines (BELs), could arise from this wind (Elvis 2000, 2003).
If instead the wind arises from the inner
edge of the obscuring torus (Krolik \& Kriss 2001) then several
disconnected gaseous components are present in AGNs, and the mass flux in the
wind is large (Netzer et al. 2003, Behar et al. 2003, Ogle et
al. 2004). A large scale wind, with a continuous radial decrease in
ionization parameter (Kinkhabwala et al. 2002) would tie together type 1
and type 2 AGNs, as high ionization kpc scale `cones' are seen around several
type 2 AGNs in Chandra images (Sako et al. 2000; Young et al. 2001;
Brinkman et al. 2002; Ogle et al. 2003), and these would produce high
ionization absorbers if viewed along the cone axis toward the nucleus.

These are quite different views of the same observations. Variability of the
X-ray absorption features can provide a direct test of these
models, as rapid variability requires clumped gas, while in continuous radial
structures of ionization parameters the variability effects should be
washed out by the average absorption in the flow. Variability of the
features can also shed light on the location of the absorber.  Studies of X-ray
absorber variability in AGNs have been undertaken for a decade or
so. Ptak et al. (1994) reported spectral variability in an ASCA
observation of NGC 3227, which could be explained by opacity changes
of ionized gas. However, this would require a decrease in ionization
state and column density of the absorber, as the luminosity of the
central source increased. On the other hand, the data was also
consistent with the variability being due to changes in the intrinsic
shape of the emitted continuum, with little change in the opacity of
the absorber. Something similar was found in two ASCA observations of MCG-6-30-15  (Reynolds
1995): the ionization parameter and column density of the absorber
were higher when the flux was lower. Using a third ASCA observation,
of  MCG-6-30-15, Otani et al. (1996) observed variability in the depth
of the o {\sc viii} edge, but no apparent change in the depth of the o
{\sc vii} edge. This led the authors to suggest that two different
regions were responsible for the absorption, and using equilibration
timescale arguments the authors were able to constrain the location of
the absorbing gas. In a ROSAT observation of NGC 4051, McHardy et
al. (1995), reported opacity changes in the absorber that did not
track linearly the changes in luminosity. More recently, and using
more comprehensive models, Nicastro et al. (1999), and Netzer et
al. (2002) have also shown variability in the opacity of the absorber,
locating the gas close to the broad emission line region.
The results from these early papers, based on low resolution
X-ray spectra, suggest variations. However these variations are hard to interpret since high
resolution X-ray spectra from Chandra and XMM-Newton now show that the 
absorbers are far more complex than the simple models used in those
papers. The high 
resolution X-ray spectra though, typically have insufficient
signal-to-noise to subdivide in time bins.

A rare exception is the integrated 900ks exposure on NGC~3783
with the Chandra  gratings. NGC 3783 (z=0.00976$\pm 0.00009$ from
stellar absorption features, de
Vaucouleurs et al. 1991) is one of the best studied Seyfert 1 galaxies
in the UV and X-ray bands, showing prominent absorption features
arising from ionized gas (e.g. Kaspi et al. 2002, Blustin et al. 2002
for the X-rays and Kraemer et al. 2001, Gabel et al 2003 for the UV).
The compiled
spectrum has given new insights on the nuclear environment of AGNs. 
There is evidence
in X-rays for absorption at radial velocities consistent with
those observed in the UV, which strongly suggests a common nature for
the outflow. Recent results by Krongold et al. (2003,
K03 hereafter) have shown that most of the features observed
in absorption (more than 100 lines and blends) can be reproduced by a simple
model  with two components, and 
that these turned out to be in pressure equilibrium. This result was
confirmed by Netzer et al. (2003) who found a third component also in
pressure balance with the other two. This third component accounts
for very highly ionized features, primarily Fe ({\sc xxiii-xxvi}).

In this paper, we present an analysis of the variability in the
opacity of the absorbers in
NGC 3783. By binning up the data into channels of size 0.25 \AA
\ (R$\sim 50$) we demonstrate
variability in the opacity of the UTA at a significance
$>10\sigma$.

\section{The High and Low state spectra of NGC 3783 \label{dat}}

NGC 3783 was observed six times using the HETGS  (Canizares et
al. 2000) on board the
{\em Chandra} X-ray Observatory (Weisskopf et al. 2000), with the Advanced
CCD Imaging
Spectrometer (ACIS, Garmire et al. 2003). The first 56 ks observation
(obsid 373) was carried out on January 2000. The following five 170 ks
observations were taken in the four months between February and
June of 2001 (see Kaspi et al. 2002, K03). In this
paper we focus on the spectra obtained by the medium energy
grating (MEG, 5-25 \AA). 

To study the temporal variability of the absorber, we proceeded
in the following way: The first two observations
(obsids 2090 and 2091) are separated by only 3 days and have extremely
similar count rates (0.38 cts s$^{-1}$, 5-25 \AA), but lower fluxes
than the rest of the observations (see Netzer et al. 2003 for detailed light
curves). Then, we coadded these two observations to obtain a
low state spectrum (hereafter LS). The observation with the highest
count rate (obsid 2093; 0.74 cts s$^{-1}$ i.e., twice the count rate
of the LS) began $\sim$31 days after the end of the LS observation. We
will refer to this observation as the ``high state'' (hereafter HS).
Our reason for choosing this grouping is that there is a difference of 16\%
in count rate between the LS with respect to obsid 2092 (0.44 cts
s$^{-1}$), and 33\% with respect to obsid 2094 (0.51 cts s$^{-1}$).
Furthermore, as shown by Netzer et al. (2003),
there is a rapid variability of the soft excess in the spectrum
of NGC 3783. Since the spectra from obsids 2092 and 2094 is also
softer, the difference in flux between these observations and the LS at
larger wavelengths ($>20$ \AA), where many of the ionization
potentials for the relevant inner shell lines are observed (including
the Fe M-shell edges responsible for the UTA), may be larger. Thus, although
coadding all these observations together would increase the
S/N of the low state, it could also wash out the opacity differences
with respect to the HS. The count rate in obsid 373 (0.60 cts
s$^{-1}$) is 20\% smaller than in obsid 2093. In addition,
the time separation between these observations (1.2
years) might introduce long term variability on the absorber because
of reasons different than flux variations (e.g. column density
variations etc.). The grouping presented here represents the best
configuration to avoid these effects.
This grouping differs from the grouping applied to the data by Netzer
et al. (2003) who considered obsid observations 2090, 2091, 2092, and
2094 as the low state and obisd observations 373 and 2093 as the high
state. The 5-25 \AA\ observed flux of the LS is $\approx
1.3\times10^{-11}$ erg cm$^2$ s$^{-1}$ (luminosity of
$2.4\times10^{42}$ erg s$^{-1}$), and that of the HS is $\approx
2.5\times10^{-11}$ erg cm$^2$ s$^{-1}$ (luminosity of
$4.5\times10^{42}$ erg s$^{-1}$).

\section{Variability of the Ionized Absorber}

\begin{figure}
\plotone{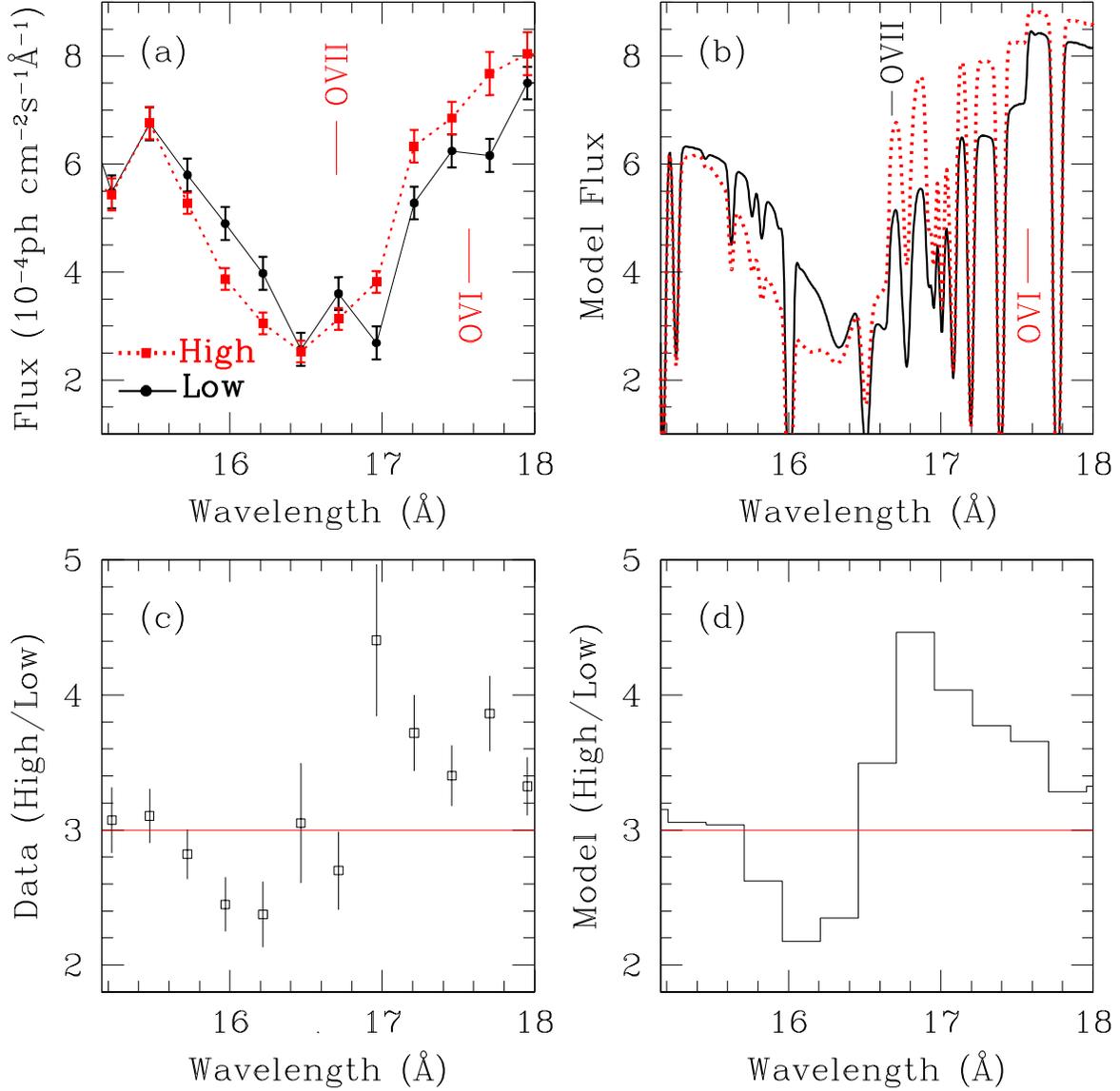} \caption[f1.eps]{(a) {\em Chandra}
HETGS spectra of NGC 3783 binned in channels of 0.25 \AA\ in
size. The low state is scaled up by a factor of 3. A clear
variation in the position of the UTA can be observed, as well as a
change in the O {\sc vi} K edge (17.6 \AA). (b) Photoionization models
for the expected opacity variation by gas in photoionization
equilibrium, as a response to a change by a factor of 2
in flux of the central source. (c) Ratio between high and low
state data. (d) Ratio between high and low state model, binned at 0.25
\AA. The significance in the variation is $>10\sigma$.
\label{spectra}}
\end{figure}

In Figure \ref{spectra}a we present a comparison of the LS and HS
spectra between 15 and 18 \AA, the region of the UTA. The
data has been grouped in bins of size 0.25 \AA\ to increase the S/N
ratio of the spectra. Although such binning washes out the signature
of isolated narrow absorption lines, it is
better for the study variability in broad blended features, such as 
the UTA. The LS has been scaled by a factor 3.0 to match
the flux level of the high state. This is the same scaling factor
applied to the data by Netzer et al. (2003), but we note that the results
presented here are not sensitive to this factor.
There is a clear shift of $\sim 0.2$ \AA\ in the position of the UTA
toward shorter wavelengths in the HS compared to the LS. This shift is
highly statistically significant and is far larger than wavelength
calibration uncertainties. The change can only be attributed to 
a physical change in the absorber.

To compare the observed opacity variation with theoretical
predictions from photoionization models, we have generated HS and a LS
absorption models with the code PHASE (K03).
We used the best (5-25 \AA) fitting model for the low ionization
component from K03 (presented in Table \ref{2ab}) as an ``average state'' of the gas producing the
UTA, as K03 analysis was based on the averaged spectrum of the six
observations. Since the count rate from the LS to the HS changed by a factor
$\sim 2$, we have selected values for the ionization parameter U of
the high and low states also separated by
a factor of 2\footnotemark.
\footnotetext{To account for any possible induced effect because of
the lack of low temperature dielectronic recombination rates (DDRs) we
have shifted the Fe ionization balance by a factor 2 with respect to
all other elements as suggested by Netzer 2004 and Kraemer, Ferland and Gabel
2004. However, the results presented here are highly independent of
the effect introduced by the lack of DRRs.}
We have used the equivalent H column density, and the turbulent and outflow
velocities of the absorber as those derived by K03 (Table \ref{2ab}).  
Figure \ref{spectra}b presents a comparison of these high and low
ionization models. There is a striking similarity
between the changes observed in the spectra and the changes predicted
by the models. In particular, the
spectra clearly shows that the shape of the UTA (along with the O
{\sc VII} K edge located at $16.7$ \AA, a feature
heavily blended
with the UTA) is varying as expected for gas close to photoionization
equilibrium. Furthermore, the O {\sc VI} K edge located at $17.6$
\AA\ also appears to be varying as predicted by the model.  


In panel (c) of Figure \ref{spectra} the ratio between the observed HS and LS
spectra is presented. The errors in this figure have been calculated
in quadrature from the errors in each spectrum. The variation in the
ratio is again highly significant, at a level $>10\sigma$ for the UTA,
and $\sim3\sigma$ for the O {\sc VI} edge.
Figure \ref{spectra}d, shows the ratio between the high state and low
state models, and for clarity we have applied the same binning factor
used for the data. Both panels show striking similarities. We conclude
that the low ionization phase (hereafter LIP, K03) responsible for the
absorption by the UTA is responding to the ionization continuum and is
possibly in photoionization equilibrium.

\begin{figure}
\plotone{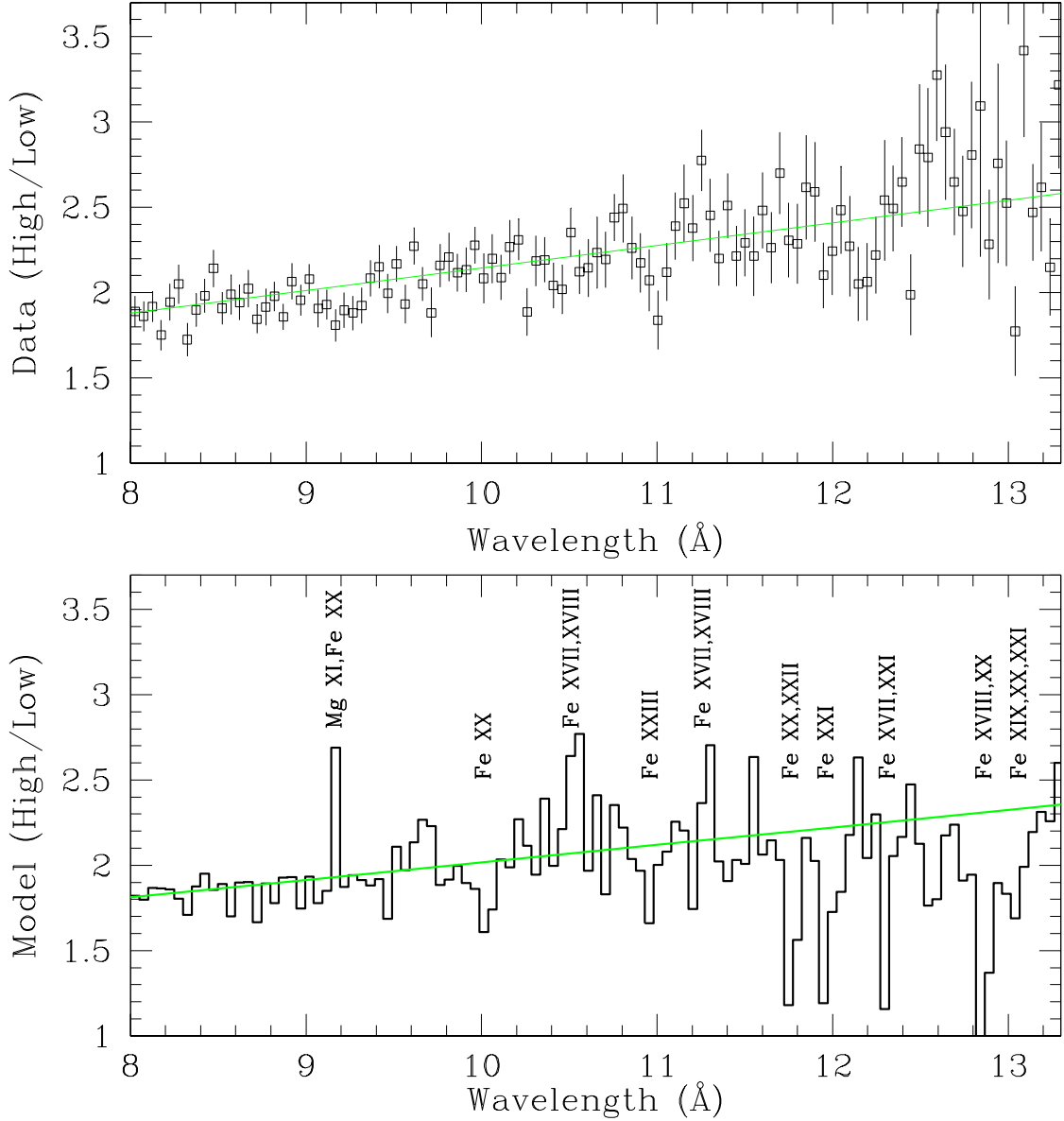} \caption[f2.eps]{Top: Ratio between
HS and LS between 8 and 13 \AA, a region dominated by absorption
produced by the HIP. Bottom: Ratio between HS and LS models
assuming changes as expected for gas in photoionization
equilibrium. Data and model are presented with bins of 0.05 \AA\ in
size. Despite the caveats of detecting variability in narrow features
(see \S \ref{lines}) it is clear that this component is not varying
linearly with the flux. 
\label{hip}}
\end{figure}

In contrast, the Fe-L shell absorption, identified by K03 as the high
ionization phase (hereafter HIP, see Table \ref{2ab}), and responsible for most of the
bound bound absorption features, does not respond linearly to the
changes in the ionizing flux, as the variations that should
be present in the HS and LS spectra are not observed. This can be seen
in the region between 8 and 13 \AA\ (see Fig. \ref{hip}), where the Fe
L-shell opacity should  have varied significantly.

Finally, we note that
the results
reported in this letter are consistent with
Figure 9 by Netzer et al. (2003), where they present the ratio between 
their choice of high and low spectra in bins of 0.1 \AA\ in size.

\section{Discussion \label{disc}}

\subsection{The Structure of Ionized Absorbers \label{struc}}

The observed variability of the UTA
helps in understanding the structure of ionized absorbers in AGN. A
popular model for the structure is that of a continuous range of
ionization structures spanning a large location in the nuclear
environment (rather than a localized absorber), and spanning more than
two orders of magnitude in ionization parameter (e.g. Ogle et
al. 2004, and references therein). In such a model many
ions of the same element have comparable column densities. Thus 
no variability is expected in the opacity of the absorber with moderate
flux variations (by a factor $< 3$),
because the average column density of each ion
remains constant in the flow. This is because the amount of
material in the line of sight changing from charge state {\em i} to
charge state {\em i+1}, as the result of a continuum variation,
is almost equally replaced by the amount of material transformed into
charge state {\em i} from charge state {\em i-1}. Thus variation is
only expected at the boundaries of the flow. In
this case the observed UTA represents the ``averaged absorption'' from
the dominant Fe charge
states in all shells  (Netzer 2004) and will remain
constant to moderate continuum variations. The observed variation
of the UTA rules then out a continuous radial range of ionization
stages, and implies that the absorbing gas is
highly clumped.

\subsection{The Low Ionization Phase}

The change in the opacity of ionized plasma
depends not only on the change of flux, but also on the change in the
spectral energy distribution (SED) of the central continuum source.
This should be kept in mind because there is a
larger change in the soft X-ray region of the spectra than in
the hard region between the LS and HS observations (Netzer et
al. 2003). However, the
MEG data does not allow us to measure spectral variations beyond
25 \AA, which is the most critical part of the spectrum for the ionization
state of the LIP. This means that the soft excess might extend up to
the Lyman limit (as assumed by Netzer et al. 2004), or might rather be
a localized feature, fading out at relatively large energies
(i.e. E$<0.5$ keV), as suggested by recent studies (Haro-Corzo et
al. 2005, in preparation). Evidently these two possibilities of SED would
produce dramatically different effects on the opacity of the ionized
gas. Additional complexity is added because at least part of the
changes observed in the soft band are not intrinsic to the continuum
source, but produced by the change in opacity of the
absorber. All these effects make an exact determination of the
expected variation of the gas 
in photoionization equilibrium impossible to achieve. Such
determination would require a self consistent global model, over time
resolved broad band data. However, we stress again, that the change in
the shape of the UTA is consistent with a
change in ionization parameter by a factor of 2. Since the observed total
flux changed also by this factor, there is support for a scenario
where the LIP is close to photoionization equilibrium.

Assuming photoionization equilibrium applies, and using the
temperature and ionization parameter deduced from the K03
model (Table \ref{2ab}), and the recombination times from the leading
charge states of Fe producing the UTA (Shull \& van Steenberg 1982),
we can set limits on the
density and location of the absorber. Considering the elapsed time
between the LS and HS ($\sim$ 31 days) as an upper limit for the destruction
(or equilibration) timescale of a single ion (see Krolik and Kriss
1995 and Nicastro et al. 1999) we find n$_e>1\times 10^4$ cm$^{-3}$
and D$<5.7$ pc.

Behar et al. (2003) reported a slight shift in the position of the UTA
in NGC 3783 during a change of flux by a factor of 2 in a timescale $\sim1.5$
days, but rejected photoionization equilibrium
arguing that the change in
the position of the UTA would have had to be 0.4 \AA\ with such change
in flux. 
However, this calculation considers only the change
in opacity for Fe.
Including absorption from all the ions in the LIP 
reduces the expected difference to 0.2 \AA, because of the presence of
bound-free transitions (particularly the O {\sc vii} and  O {\sc vi}
K edges, see Fig. \ref{spectra}b).
Therefore, the variation observed by
Behar et al. (2003) might still be consistent with photoionization
equilibrium. Furthermore, the S/N ratio of their low and high state
data might not be enough to detect such variation in a significant
way. Deciding this
is beyond the scope of the present paper, however we note the
following: (1) If the XMM S/N ratio is indeed sufficient to detect the
expected opacity variation, and this variation is not observed, then
the gas should be located
between 5.7 and 1.3 pc from the central source
 excluding the Broad Line
Region, as suggested by Behar et al. (2003)\footnotemark.
(2) If the expected variation is detected, the gas should be
closer than 1.3 pc. (3) If the S/N
ratio is not sufficient to detect the expected opacity
variation, then the gas can lie anywhere within 5.7 pc.
\footnotetext{Note the different upper limit with
respect to Behar et
al. 2003 (D$>$2.8 pc). For consistency, we have used in our calculation the
equilibration (or destruction) time
(which is inversely proportional to both the photoionization rate PI
of ion {\em i} into ion {\em i+1} and the recombination rate of ion
{\em i} into ion {\em i-1}), while Behar et al. used only the inverse
of the PI rate, which is only an upper limit to the equilibration
time. Then, the upper limit imposed here should be a more
realistic measurement.}

In these last two cases the
location of the absorber could thus, well be at
subparsec scales, as suggested by Reeves et al. (2004, Distance $<$
0.1 pc) for Fe
K-shell (the third and hottest absorption component in NGC
3783). This is also consistent with measurements for ionized absorbers
in other objects, which suggest that the location of the absorber
might be similar to that of the broad emission line region, e.g. NGC 4051
(Nicastro et al. 1999), NGC 3516 (Netzer et al. 2002), and NGC 7469 (Kriss
et al. 2003, Blustin et al. 2003). 

In any case, the detected
variability presented here rules out the presence of an
absorber extending tens or hundreds of parsecs outside the nucleus (and
a continuous range of ionization stages, see \S \ref{struc}),
and rather suggests a localized absorber, as has also been suggested by the
evidence of transverse motion of the absorbing gas with respect to our
line of sight in several objects (see Crenshaw, Kraemer and George
2003, and references therein). Given these properties, it is clear
that the absorber (at least the LIP) in the nuclear environment of NGC
3783 is very different from the extended X-ray emission gas observed is
Seyfert 2 galaxies.

\subsubsection{Using the UTA to Constrain the Changes in the
  Ionization State of the Gas \label{lines}}

It is important to note that the UTA sets tight constraints in the
ionization state of the absorber (K03) because (1) this feature is
strongly dependent on the charge states of Fe producing it, since
several charge states of the same element are observed at the same
time, and (2) because it is a broad feature, and therefore its
detection involves a
large number of detector channels which allows for a heavy binning of
the spectra to consider large number of counts (see also Behar et
al. 2003). We stress, then, that the observed change in the position
of the UTA is a highly robust measurement of the opacity variation in
the ionized absorber in NGC 3783.

\subsection{The High Ionization Phase}
 
In contrast to the LIP, the HIP does not vary as expected in
photoionization equilibrium. However, since
there are no very broad features in this component (like for instance
the UTA in the LIP), and
many features are composed by
several blends of very different
charge states from different elements, it is
not even clear whether this component is changing as a whole or not, or
even to establish if the ionization state is increasing or
decreasing. Such behavior might have several explanations: 
\newline\noindent (1)
This component may have such a low density, that the timescale for
the gas to reach photoionization equilibrium might be larger than the
variability timescale of NGC 3783. In this situation the gas would be
out of equilibrium, and, rather than changing as a whole,  each
element would vary according to the recombination timescales of
its dominant charge states. Since this timescale varies from ion to
ion and element to element, the opacity of the gas would not change as
a whole, but rather some elements might seem to get more ionized,
while other may seem to decrease in state of ionization (see Nicastro
et al. 1999). This would imply n$_e<1\times 10^5$  cm$^{-3}$ and
D$>$0.6 pc (Netzer et al. 2003).  
\newline\noindent (2) The second possibility is that the opacity is varying
very little
because the gas lies in the intermediate stable branch
of the thermal equilibrium T vs. U/T  curve (or S-curve). The length of this
branch strongly
depends on the metallicity of the gas, and while it can be small for
solar metallicity, it can be much more extended for values 3
times solar (see Komossa and Fink 1997). Even
the 900 ks exposure of NGC 3783 is insensitive to deviations by these
factors in
metallicity (K03). Therefore, if the metallicity of
this component is close to solar, the HIP might not be changing
simply because it is confined to the intermediate stable branch of the
thermal equilibrium curve (K03). 
\newline\noindent (3) Finally, it is also possible that the opacity is
not changing as
expected because the gas is pressure confined (in this case by a third
hotter component, as suggested by Netzer et al. 2003). As has been amply
discussed by Krongold et al. (2004), in a case of a 3-phase confined
medium, the confined components have to follow the confining medium on
the thermal equilibrium curve. In such a scenario, while the variation
of the LIP is expected to be similar to the one observed in
photoionization equilibrium, because of the shape of the S-curve, the
variability of the HIP might be completely different,
even decreasing in ionization while the flux increases (see Krongold
et al. 2004 for a full explanation). In this case this gas must, of
course, share the same physical location as the LIP and of the third,
hottest component.

Deciding among these possibilities would require more data, and
models including time evolving photoionization effects.

\section{Conclusions}

Using the 900 ks {\em Chandra} HETGS spectrum of NGC 3783, we
unambiguously ($> 10\sigma$) detect opacity variations in the
Fe M-shell unresolved
transition array (UTA), in response to an ionizing continuum change by
a factor of 2. The UTA is the best diagnostic for detecting
opacity changes in ionized plasma. The changes are as expected for gas
close to photoionization equilibrium implying a density n$_e >1\times
10^4$ cm$^{-3}$. This result locates the
absorbing material within
6 pc of the central source and are consistent with estimates by
Reeves et al. (2004) for another absorbing
component. The strong change in ionization state rules out a picture
where the absorber extends radially in a continuous flow over 
hundreds of parsecs, implying that this gas is different than the one
observed in emission in Seyfert 2 galaxies.
The observations also imply that the gas is heavily clumped, since in
a continuous distribution of ionization stages no variability is
expected with moderate flux variations.

\acknowledgements 

This research has been partly supported by NASA
Contract NAS8-39073 ({\em Chandra} X-ray Center), NASA grant NAS G02-3122A,
and Chandra General Observer Program TM3-4006A.

\clearpage

\begin{deluxetable}{lcc}
\tablecolumns{3} \tablewidth{0pc} \tablecaption{Two Phase Absorber
Model for the full (900 ks) {\em Chandra} observation of  NGC~3783
(Krongold et al. 2003) \label{2ab}}

\tablehead{\colhead{Parameter} & \colhead{High-Ionization} &
\colhead{Low-Ionization} } \startdata
Log U$^a$ &  0.76$\pm0.1$ & -0.78$\pm0.13$ \\
Log N$_{H}$ (cm$^{-2}$)$^a$ &  22.20$\pm.22$ & 21.61$\pm0.14$ \\
V$_{Turb}$ (km s$^{-1}$)& 300 & 300 \\
V$_{Out}$ (km s$^{-1}$)$^a$& 788$\pm138$ & 750$\pm138$ \\
T (K)$^b$& $9.52\pm0.44\times10^5$ & $2.58\pm0.39\times10^4$ \\
\enddata
\tablenotetext{a}{Free parameters of the model.}
\tablenotetext{b}{Derived from the column density and ionization
parameter, assuming photoionization equilibrium.}
\end{deluxetable}

\clearpage

\clearpage

\end{document}